# A study on implementing a multithreaded version of the SIRENE detector simulation software for high energy neutrinos

Petros Giannakopoulos[2], Michail Gkoumas[2], Ioannis Diplas[2], Georgios Voularinos[2], Theofanis Vlachos[1], Konstantia Balasi[1], Ekaterini Tzamariudaki[1], Christos Filippidis[1,2], Yiannis Cotronis[2], Christos Markou[1]

[1]NCSR Demokritos
[2]National and Kapodistrian University of Athens

**Abstract.** The primary objective of SIRENE is to simulate the response to neutrino events of any type of high energy neutrino telescope. Additionally, it implements different geometries for a neutrino detector and different configurations and characteristics of photo-multiplier tubes (PMTs) inside the optical modules of the detector through a library of C++ classes. This could be considered a massive statistical analysis of photo-electrons. Aim of this work is the development of a multithreaded version of the SIRENE detector simulation software for high energy neutrinos. This approach allows utilization of multiple CPU cores leading to a potentially significant decrease in the required execution time compared to the sequential code. We are making use of the OpenMP framework for the production of multithreaded code running on the CPU. Finally, we analyze the feasibility of a GPU-accelerated implementation.

## 1 Introduction

SIRENE [1] receives as input an event file, ROOT [2] formatted, with the information from the generator level and uses a separate file for the description of the detector geometry. The Probability Density Functions of the arrival time of light (PDFs) [3] are read from 4 files, each containing the data of the Cumulative Density Functions (CDFs) of direct light from a muon, scattered light from a muon, direct light from an Electro-Magnetic shower and scattered light from an Electro-Magnetic shower, respectively. The processing steps followed by SIRENE are outlined as follows: 1) Parse Event from input file, 2) Propagate muon(s) by simulating energy loss and EM-showers and generating hits originating from direct and single scattered light, 3) Process shower particles from primary vertex and generate hits from direct and single scattered light, 4) Merge hits to speed-up TriggerEfficiency, 5) Write computed event to ROOT formatted output file.

## 2 Overview of SIRENE

SIRENE implements a number of nested "for" loops for the computation of hits on OMs and on the PMTs within each OM: 1) Event "for" loop: consists of a number of particles described as Tracks.

Events are independent of each other. 2) Tracks "for" loop: Describes the path of each particle, starting with an initial energy E0 and initial position on z-axis z0. Energy loss and propagation is calculated in steps on each iteration. Hits from both direct/scattered light from muons/EM showers are computed. Hits registered on PMTs along the particle's path are computed via two procedures: a) Integration of the CDF via polynomial interpolation approximation for calculating photon emissions, b) Poisson Distribution applied on arrival times of photo-electrons for calculating the number of photo-electrons on a Module and included PMTs.

## 2.1 Time requirements

The most time consuming functions of SIRENE, in descending order of share of the total execution time, are: 1) The calculations for hit probabilities on OMs/PMTs for each track and 2) The particle propagation which includes the coordinate system transformations for the Tracks and PMTs. Decomposing, 1) accounts for ~90% of total execution time, while 2) and various secondary functions, such as I/O, amount to the rest 10%. It is evident that the calculations of hit probabilities are by far the heaviest part of SIRENE, so speeding up its execution could have a significant positive effect on the total runtime.

## 3 Multithreaded SIRENE

Events are independent of each other, do not exchange any data and a study requires the amount of approximately one million events or higher. Therefore they can be spread over multiple compute nodes by sending a number of events for computation to each node. This can be trivially done via the use of batch jobs. Each node will produce an output for the events assigned to it. When all nodes are finished we can collect and sort the outputs to piece together the complete simulation. The main focus of our implementation is how to take advantage of the multiple cores available in the CPU(s) within each node in order to accelerate the computations pertaining to each event. By examining SIRENE's most time consuming segments of the original sequential code, the next step is to identify the most suitable methods of parallelization for each of these segments:

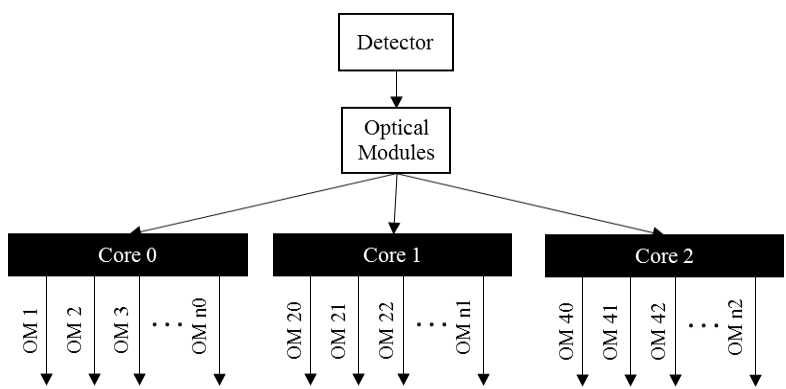

**Figure 1.** OMs spread over multiple CPU cores

a. Hit detection

For each particle track all the calculations for probabilities of a hit are performed on each OM of the detector. For each OM, the probabilities of this track reaching the OM and registering hit(s) on the PMT(s) inside are calculated. There are typically hundreds of OMs in a detector providing enough work that can be shared between many CPU cores. A team of OMs can be assigned to each core allowing a potentially significant speedup over having all OMs assigned to a single core. The team's size can vary from 1 to N, where N the total number of OMs in the detector, depending on the available CPU cores. This is visualized in Figure 1.

b. Coordinate system rotations

Energy loss and position are calculated in discrete steps along the particle's path. To simplify the simulation, a particle's coordinate system is rotated such as the path lies only on the z-axis. The same transformation is applied to each OM of the detector so that the enclosed PMTs point along the z-axis. Since this rotation is applied to every OM of the detector it can also be executed in parallel by assigning a team of OMs to each thread.

c. PDFs to CDFs conversion

Conversion of PDFs to CDFs doesn't require a significant amount of time, but the construction of the CDF tables can also be parallelized for additional speedup.

## 3.1 Implementation

Analyzing SIRENE, as shown in Hit Detection, emerged the need for the use of more CPU cores. Another key point is the fact that the output is saved in a shared object, Event, which is ROOT formatted. As a result, one 'machine' should be involved, otherwise multiple 'machines' would require a great deal of overhead. OpenMP [4] is able to operate under these circumstances. It is implemented by enlisting one thread per OM. Each thread is capable of creating sub-threads as needed, according to the inner loops and the available system resources. The hit calculations of muon and EM showers are totally independent of each other, in different consecutive blocks inside different OM blocks. For the implementation, the appropriate OpenMP constructs are employed for dividing the work to several threads and for conflict avoidance when shared variables are accessed from multiple threads. Alterations to the original code structure are made wherever needed in order to facilitate more efficient parallel execution via optimizing order of execution and variable accesses for performance.

## 3.2 Performance results

Comparisons between the sequential and the parallel implementations of SIRENE were performed to gauge the benefits in performance offered by parallelization (Figure 2). On our test system, the part of SIRENE responsible for detection of photon hits on PMTs required ~10 ms per event for the sequential version and ~4 ms for the multithreaded version across all particle energies. We can observe that with 4 cores (= 4 threads) available to the multithreaded version, hit detection on PMTs needs ~2.5x less time to complete compared to the sequential version running on only 1 core (= 1 thread). Total execution time measures ~11 ms for the sequential version and ~5.5 ms for the multithreaded version, a ~2.1x decrease in runtime. Total runtime includes non-parallelizable functions, such as I/O, which impose a limit to the max speedup that can be achieved via parallelization. These functions make up only a small part of SIRENE so their negative impact on total speedup remains small yet measurable.

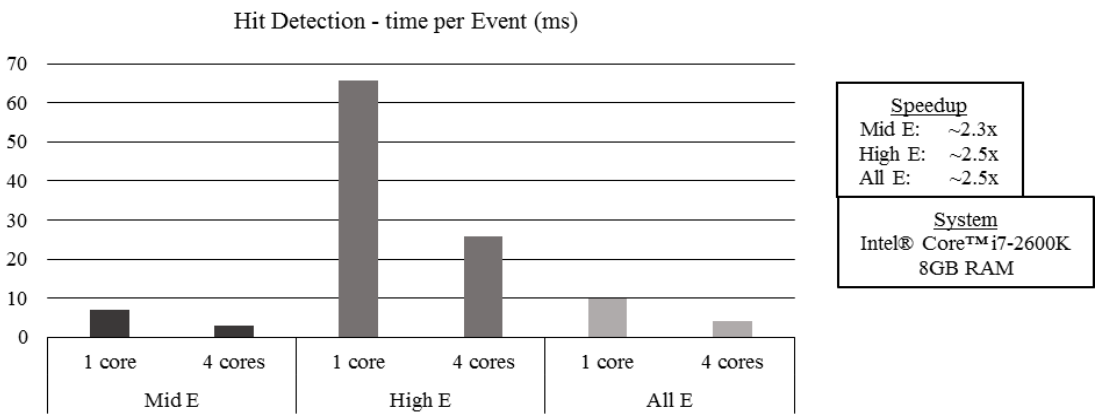

**Figure 2.** Runtimes of sequential and multithreaded versions of SIRENE

By dividing the whole energy spectrum (All E) into three slices (Low E – 100GeV to 10TeV, Mid E – 10TeV to 1PeV, High E – higher than 1PeV), we can observe the contribution of each. In this case, most particles have low E, some mid E and few have high E. The higher the energy, the more photons are produced and the higher the computational load, which is evident by the runtimes of mid and high E events in Figure 2. Low E events are not included because of their very short runtime, yet due to their large number, they have the highest contribution to the average runtime across all energies. We can also observe a bit higher speedup for the High E's (~2.5x) compared to Mid E's (~2.3x) and Low E's (~2x) which can be accounted to the increasing computational load.

In order to test the scaling of the implementation with the number of threads, we also measured runtime using 2 cores. We observed ~1.7x speedup with 2 threads compared to ~2.5x with 4 threads. This points to factors that limit the scaling of the implementation with the available threads. The most limiting factor is thread idling created by the computational load imbalance between different loop iterations on account of varying particle energies. Measures have been taken to mitigate the impact of load imbalance, such as using dynamic thread scheduling, however these methods can incur additional overhead.

## 4 Conclusion

SIRENE multithreaded implementation has delivered a positive result in terms of time saving by taking advantage of the multiple cores present in every modern CPU. However, it is possible that even higher gains can be achieved by utilising the compute capabilities of GPUs. We have already implemented a GPU version of the coordinate systems transformations [5] by employing the CUDA framework. The relatively small computational workload provided by 1 event constrains the speedup to ~4x and GPU is underutilized. In order to get the maximum of the GPU's capabilities a parallel algorithm is needed to address more data inputs in the bounds of system's memory resources. Proof is the performance result of the GPU implementation. However, this can change when the hit detection computations, which are significantly heavier, are also offloaded to the GPU. In order to achieve this, significant functions which SIRENE uses repeatedly, such as Random Number Generation and Polynomial or Spline Interpolation, need to be implemented using parallel versions of these functions, suitable to be ran on a GPU. For example, the NVIDIA CUDA Random Number Generation library – cuRAND [6] can be used for fast random number generation on the GPU and the Cubic B-Spline interpolation algorithm for CUDA [7] can replace the CPU-oriented algorithm.